\documentclass[%
 reprint,
superscriptaddress,
 amsmath,amssymb,
 aps,
prb,
]{revtex4-1}

\usepackage{graphicx}
\usepackage{dcolumn}
\usepackage{bm}

\usepackage{color}

\begin{document}

\preprint{APS/123-QED}

\title{Boson peak dynamics of glassy glucose studied \\by integrated terahertz-band spectroscopy}

\author{Mikitoshi Kabeya}
\author{Tatsuya Mori}%
 \email{mori@ims.tsukuba.ac.jp}
\affiliation{%
 Division of Materials Sciences, University of Tsukuba, 1-1-1 Tennodai, Tsukuba, Ibaraki 305-8573, Japan
}%

\author{Yasuhiro Fujii}
\author{Akitoshi Koreeda}%
\affiliation{%
 Department of Physical Sciences, Ritsumeikan University, 1-1-1 Noji-higashi, Kusatsu, Shiga 525-8577, Japan
}%

\author{Byoung Wan Lee}
\author{Jae-Hyeon Ko}
\affiliation{
 Department of Physics, Hallym University, 1 Hallymdaehakgil, Chuncheon, Gangwondo 24252, Korea
}%

\author{Seiji Kojima}
\affiliation{%
 Division of Materials Sciences, University of Tsukuba, 1-1-1 Tennodai, Tsukuba, Ibaraki 305-8573, Japan
}%

\date{\today}

\begin{abstract}
We performed terahertz time-domain spectroscopy, low-frequency Raman scattering, and Brillouin light scattering on vitreous glucose to investigate the universal boson peak (BP) dynamics.
In the spectra of $\alpha(\nu)/\nu^{2}$ ($\alpha(\nu)$: absorption coefficient), the BP is clearly observed around 1.1 THz.
Correspondingly, the complex dielectric constant spectra show a universal resonance-like behavior only below the BP frequency.
As a novel analytical scheme, we propose the relative light-vibration coupling coefficient (RCC), which is obtainable from the combination of the far-infrared and Raman spectra.
The RCC reveals that the infrared light-vibration coupling coefficient $C_{\rm IR}(\nu)$ of the vitreous glucose behaves linear on frequency which deviates from the Taraskin's model of $C_{\rm IR}(\nu)=A+B\nu^{2}$ [Phys. Rev. Lett. $\bf 97$, 055504 (2006)]. The linearity of the $C_{\rm IR}(\nu)$ might require modification of the second term of the model. The measured transverse sound velocity shows an apparent discontinuity with the flattened mode observed in the inelastic neutron scattering study [Phys. Rev. B $\bf 85$, 134204 (2012)], and suggests a coupling between the transverse acoustic and flattened modes.
\begin{description}
\item[PACS numbers]
61.43.Fs, 63.50.-x, 78.47.-p
\end{description}
\end{abstract}

\pacs{Valid PACS appear here}
\maketitle


\section{\label{sec:level1}Introduction}

The boson peak (BP) is a universal feature of glassy materials observed in the terahertz (THz) region, and it is often regarded as ``the excess vibrational density of states'' $g(\nu)$ (VDOS)\cite{Nakayama2002}.
The universal feature is recognized as a peak in the spectrum of $g(\nu)/\nu^{2}$, indicating a deviation from the Debye model of the crystalline acoustic phonon system.
The BP in structural glasses is detectable by inelastic neutron (INS) or X-ray (IXS) scattering, low-frequency Raman scattering (LFRS), the low temperature thermal properties, and  also by {\it far-infrared} (IR) {\it spectroscopy} \cite{Matsuishi1986, Hutt1989, Ohsaka1994, Ohsaka1998}.
Numerous studies has been done regarding the BP dynamics, however, the microscopic origin is not yet fully understood.
Furthermore, there have been almost no recent terahertz (THz) spectroscopic studies of the BP dynamics reported \cite{Naftaly2005, Taraskin2006, Shibata2015, Kojima2015, Sibik2016}, in spite of advanced development of the  technique of the THz light source and detection.

To elucidate the origin of the BP of glassy materials, significant experimental and theoretical studies have been done over the past several decades.
As the experimental studies, INS\cite{Buchenau1986, Yamamuro2000, Nakamura2001,  Orsingher2010PRB, Violini2012, Zanatta2013} and IXS\cite{Baldi2010, Baldi2011, Ruta2012} are the most powerful methods, and an acoustic phonon-like behavior and flattened dispersion curves have been observed in many glasses \cite{Nakamura2001, Orsingher2010PRB, Violini2012, Zanatta2013}.
Meanwhile, the LFRS technique can easily access the BP of not only inorganic glasses, but also organic glass formers due to the unnecessary deuteration\cite{Phillips1981, Nemanich1977, Malinovsky1986, Kojima1993, Surovtsev2002}.
Regarding the thermal properties, the low temperature specific heat $C_{p}$ \cite{Zeller1971, Phillips1981, Baldi2016, Carini2016} shows the BP in $C_{p}/T^{3}$, and the thermal conductivity shows a plateau in the corresponding temperature region\cite{Zeller1971, Phillips1981}.
Far-IR spectroscopy can also detect the BP and several studies have been reported using a high-pressure mercury lamp\cite{Hutt1989, Ohsaka1998, Matsuishi1986, Ohsaka1994}.

As theoretical approaches, there are several candidates, such as the phonon-saddle transition \cite{Grigera2003} based on the Euclidean random matrix theory \cite{Grigera2002}, an approach from the mode coupling theory \cite{Gotze2000}, the soft potential model \cite{Parshin1993, Parshin2003, Parshin2007}, the localized anharmonic modes of clusters \cite{Duval1990}, the hybridization of the local and acoustic modes \cite{Nakayama1998PRL, Klinger2001}, and attribution to the transverse acoustic (TA) van Hove singularity of the corresponding crystal \cite{Schirmacher1998, Taraskin1999, Taraskin2001, Pilla2004}.

A recent important study of the inelastic nuclear scattering for a sodium-iron silicate glass, Na$_2$FeSi$_3$O$_{8.5}$, reported by Chumakov $et$ $al$.\cite{Chumakov2011} revealed that the gradual transformation of the BP to the TA van Hove singularity of the crystalline state occurs.
Other INS studies of a densified silica glass\cite{Chumakov2014} and amorphous iron\cite{Chumakov2015} have also confirmed their approach.

As already mentioned, past researchers of far-IR spectroscopy have well recognized that the BP is detectable by the IR method\cite{Matsuishi1986, Hutt1989, Ohsaka1994, Ohsaka1998}, however, recent terahertz spectroscopy researchers almost missed it and did not succeed in detecting the BP, especially, by terahertz time-domain spectroscopy (THz-TDS). Recent researchers have regarded the BP as the broad peak appearing in the imaginary part of the complex dielectric constant $\varepsilon''(\nu)$ in the THz region \cite{Loidl2012, Kojima2003, Mori2015}, although the name of the broad peak has recently been changed to the ``VDOS peak'' \cite{Parrott2015}. In order to understand how the BP appears in the IR spectra, we must consider the following relation derived from the linear response theory for disordered materials \cite{Galeener1978},
\begin{equation}
\alpha(\nu) = C_{\rm IR}(\nu)g(\nu),
\label{eq-CIR}
\end{equation}
where $\alpha(\nu)$ and $C_{\rm IR}(\nu)$ are the absorption coefficient and IR light-vibration coupling coefficient, respectively \cite{footnote1}.
We re-note that the BP appears in the $g(\nu)/\nu^{2}$ spectra indicating a deviation from the Debye model for the three-dimensional phonon system with a linear dispersion relation, and re-express Eq. (\ref{eq-CIR}) as follows:
\begin{equation}
\frac{\alpha(\nu)}{\nu^{2}} = C_{\rm IR}(\nu)\frac{g(\nu)}{\nu^{2}}.
\label{eq-af2}
\end{equation}
We can then understand that the BP in the IR spectra will appear in the representation of $\alpha(\nu)/\nu^{2}$, through $C_{\rm IR}(\nu)$.
Naftaly $et$ $al$. \cite{Naftaly2005} had recognized this and they represented the BP as $\alpha(\nu)/\nu^{2}$, although this recognition has not spread to other researchers in the fields of both glass systems and THz spectroscopy.
Recently, we pointed out that the BP in the sense of $\alpha(\nu)/\nu^{2}$ is detectable by THz-TDS for the pharmaceutical indomethacin (IMC) glass \cite{Shibata2015, Kojima2015}, which is one of the organic glass formers, then some researchers noticed it \cite{Sibik2016} or re-recognized the potential of the THz-TDS for detection of the BP.
Consequently, THz-TDS is a suitable technique to detect the BP. As we will describe in this paper, the BP in the IR spectra is derived from a universal and characteristic structure of the complex dielectric constants.

Vitreous glucose is one of hydrogen-bonded molecular glass-formers and a model substance for study of the BP dynamics, because the BP frequency locates at approximately 1 THz \cite{Violini2012}, which is  suitable for the THz-TDS having a detector with a photo-conductive antenna \cite{Shibata2015, Helal2015}.
In addition, the $T_{g}$ of vitreous glucose is about 310 K \cite{Wungtanagorn2001}, slightly higher than room temperature, therefore, we can perform the THz-TDS measurement without a liquid cell, resulting in a high quality complex dielectric spectra.

The structure of the glassy state of glucose has been investigated by neutron scattering and NMR \cite{Tromp1997} and the results revealed that the vitreous glucose includes both $\alpha$- and $\beta$-D-glucose molecules \cite{Brown1979, Chu1968} at the ratio of about 1 to 1.
Macroscopic investigations have been done regarding its density\cite{Parks1928}, thermal expansion\cite{Parks1934}, viscosity\cite{Parks1934, Ollett1990}, and fragility\cite{Angel1994}.
In a broadband dielectric spectroscopic study \cite{Chan1986, Nole1992, Kaminski2006} of the liquid-glass transition, the $\alpha$-relaxation, JG-$\beta$-relaxation assigned as the Johari-Goldstein secondary relaxation, and $\gamma$-relaxation possibly originating from the hydrogen-bonding scheme, have been investigated \cite{Kaminski2006}.
Based on the THz-TDS for the glassy state, Walther $et$ $al$. observed a broad absorption toward the higher frequency in the $\alpha(\nu)$ spectra of the THz region \cite{Walther2003}, although they were not aware of the existence of BP in their spectra.
Furthermore, the observation of the BP in vitreous glucose by LFRS  has not yet been reported to the best of our knowledge.
From the INS study of Violini \cite{Violini2012}, an almost flat $L$-mode and a dispersive $H$-mode were detected in the dynamic structure factor $S(Q,\omega)$ at about 6 meV and 20 meV, respectively, and the relation with the acoustic modes has been discussed. A blurred and unclear BP in $g(\nu)/\nu^{2}$ has been observed at around 3 meV at room temperature, and the BP energy is lower than the $L$-mode energy.

In this study, we performed THz-TDS on vitreous glucose and detected the BP in the $\alpha(\nu)/\nu^{2}$ spectra, and discuss the dynamics of the complex dielectric constants.
We also perform LFRS to detect the BP and compare the results with the THz-TDS.
In addition, as a novel approach to analyze the BP dynamics by IR and Raman spectroscopies, we propose a relative light-vibration coupling coefficient which is obtained from the results of the combination of the THz-TDS and LFRS.
We also demonstrate that the approach is advantageous for evaluation of the coupling coefficient without the VDOS spectra.
Finally, we determine the transverse sound velocity from the Brillouin light scattering, and will find an anomaly of the discontinuity between the TA mode and flattened $L$-mode observed in the previous INS study \cite{Violini2012}.

\section{Experimental}

D-(+)-Glucose, having the melting temperature of 423~K\cite{Hurtta2004}, was purchased from Sigma-Aldrich, Inc. The glassy states of glucose were prepared by melt-quenching of the crystalline powdered glucose from 473~K to room temperature of 293~K, which is below the glass transition temperature of 310~K\cite{Wungtanagorn2001}, under normal atmospheric conditions. To avoid caramelization, the times in the liquid state were kept as short as possible, and the obtained samples were nearly colorless, transparent glasses. The sample for THz-TDS is a disk-shaped pellet with the diameter of about 15~mm and thickness of 1.035~mm. All the samples used for the measurements were prepared by the same procedure as much as possible.

THz-TDS with the low temperature-grown GaAs photoconductive antennas for both the emitter and detector (RT-10000, Tochigi Nikon Co., Ltd.), covering the frequency range of 0.25 - 2.25~THz, were carried out using the standard transmission configuration for the temperature dependent measurements. The temperature was varied from 14~K to 320~K using a liquid helium flow cryostat system (Helitran LT-3B, Advanced Research Systems, Inc.) \cite{Igawa2014, Helal2015, Shibata2015}. We employed the Cherenkov-type MgO doped LiNbO$_3$ emitter (TAS7500SU, Advantest Corp.) \cite{Igawa2014, Mori2014IOP, Kojima2014IOP, Shibata2015} for the room temperature broadband measurement.

Confocal micro-Raman measurements were performed with a depolarized backscattering geometry \cite{Fujii2016}. A frequency doubled diode-pumped solid state (DPSS)  Nd:yttrium-aluminum-garnet laser oscillating in a single longitudinal mode at 532~nm (Oxxius LMX-300S) was employed as the excitation source. A home-built microscope with  ultra-narrowband  notch  filters  (OptiGrate) was used for focusing the excitation laser and collecting the Raman-scattered  light. The scattered light was analyzed by a single monochromator  (Jovin-Yvon, HR320, 1200~grooves/mm) equipped with a charge-coupled-device (CCD) camera (Andor, DU420). 

The Brillouin scattering spectrum was measured using a conventional tandem 6-pass Fabry-Perot interferometer (TFP-1, JRS Co.) \cite{Ko2013}. An ordinary photon counting system and a multi-channel analyzer were used to accumulate the output signals. A DPSS laser (Excelsior 532-300, SpectraPhysics) at the wavelength of 532~nm was used as the excitation source. A forward, symmetric scattering geometry with the angle of 90$^\circ$ was used to measure the sound velocity without knowing the refractive index.

\section{Results and discussion}

\subsection{Boson peak detection by THz-TDS}

\begin{figure}[b]
\includegraphics[width=7cm]{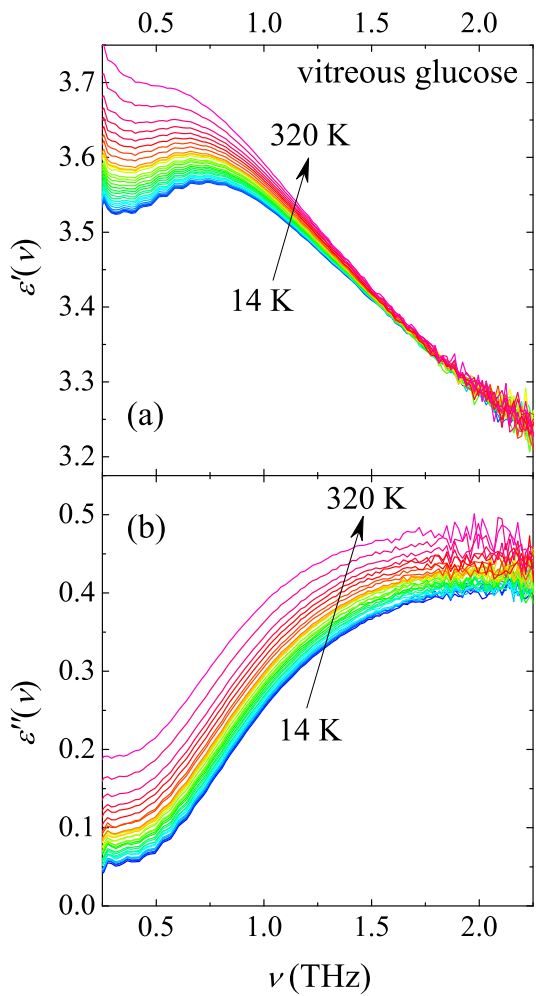}
\caption{(Color online) Temperature dependence of (a) the real and (b) imaginary part of the complex dielectric constants of the vitreous glucose during the heating process, at 14, 20, $\cdots$, 320~K. The data are plotted every 10~K above 20~K. For convenience, 1~THz = 33.3 ~cm$^{-1}$ = 48~K = 4.14~meV.}
\label{fig1}
\end{figure}

Figure~\ref{fig1} shows the temperature dependence of the complex dielectric constant $\varepsilon(\nu)$ of the vitreous glucose during the heating process measured by the THz-TDS. At the lowest temperature of 14~K, a {\it resonance-like} line shape structure was observed in both the real $\varepsilon'(\nu)$ and imaginary $\varepsilon''(\nu)$ parts of the $\varepsilon(\nu)$. This resonance-like structure deviates from both the damped harmonic oscillator (DHO) model and the Debye relaxation model, and it is the universal feature of the response function of the glassy materials in the THz region. These behaviors have not been pointed out or assumed as experimental error in the earlier studies of glassy materials by THz-TDS \cite{Grischkowsky1990, Walther2003}.
``The resonance-like'' in the spectra  means that the DHO model-like behavior appears only below about 1~THz, while above this frequency, the spectra become broad and the resonance-like feature disappears. We note that the crossover frequency exactly corresponds to the BP frequency in the IR spectra.
Regarding the temperature dependence, with the increasing temperature during the heating process, the values of $\varepsilon'(\nu)$ and imaginary $\varepsilon''(\nu)$ in the low frequency region increase and the resonance-like structure becomes blurred, leading to softening of the peak position of $\varepsilon'(\nu)$.
These behaviors are caused by the appearance of several relaxation processes in the low frequency region relating the liquid-glass transition, i.e., the JG-$\beta$ and $\gamma$ relaxations influence below the $T_g$, and the $\alpha$-relaxation above the $T_g$ \cite{Kaminski2006}. The excess wings of the relaxation processes detected by THz-TDS have been discussed for the hydrogen-bonded glass formers \cite{Sibik2013, Sibik2014, Parrott2015}.

\begin{figure}
\includegraphics[width=7cm]{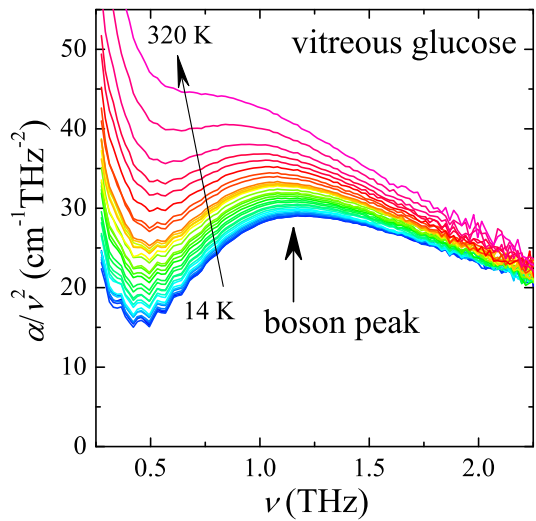}
\caption{(Color online) Temperature dependence of boson peak plot $\alpha(\nu)/\nu^{2}$ of the vitreous glucose during the heating process, at 14, 20, $\cdots$, 320~K. The data are plotted every 10~K above 20~K. For convenience, $\alpha(\nu) = \nu \varepsilon''(\nu) \cdot 2\pi/(cn'(\nu))$, where $c$ and $n'(\nu)$ are the speed of light and the refractive index, respectively.}
\label{fig2}
\end{figure}

Figure~\ref{fig2} shows the spectra of $\alpha(\nu)/\nu^{2}$ that represent the BP of the IR spectra.
At the lowest temperature of 14~K, a peak is clearly observed at 1.17~THz and this is the BP in the IR spectrum.
The results of the selected temperatures are listed in Table \ref{table1}.
The BP shifts toward a low frequency as the temperature increases, and the line shape becomes broad.
This is mainly caused by the excess wings of the several relaxation modes seen in the $\varepsilon''(\nu)$ spectra and might not be an intrinsic effect of the BP behavior.
We can now recognize that the resonant-like behavior of the $\varepsilon(\nu)$ shown in Fig.~\ref{fig1} appears only below the BP frequency, $\nu_{\rm BP-IR}$.

It is summarized that the universal behaviors of the THz spectra of the glassy materials are the appearances of a resonance-like line shape of $\varepsilon(\nu)$ below the $\nu_{\rm BP-IR}$ and a peak in the $\alpha(\nu)/\nu^{2}$ spectra, not in the $\alpha(\nu)$ or $\varepsilon''(\nu)$ spectra.
It is worth noting that, even at 14~K and below 0.5~THz, $\alpha(\nu)/\nu^{2}$ increases toward the low frequency. The behavior might be related to the nearly constant loss which is observed in the dielectric constant measurements \cite{Sidebottom2002} where the relevance to the two level system \cite{Phillips1972} is discussed. Although it goes beyond this study, the THz-TDS can access such a low frequency and a low temperature universal excitation of the glass system.

\subsection{Boson peak spectra in Raman scattering}

\begin{figure}
\includegraphics[width=7cm]{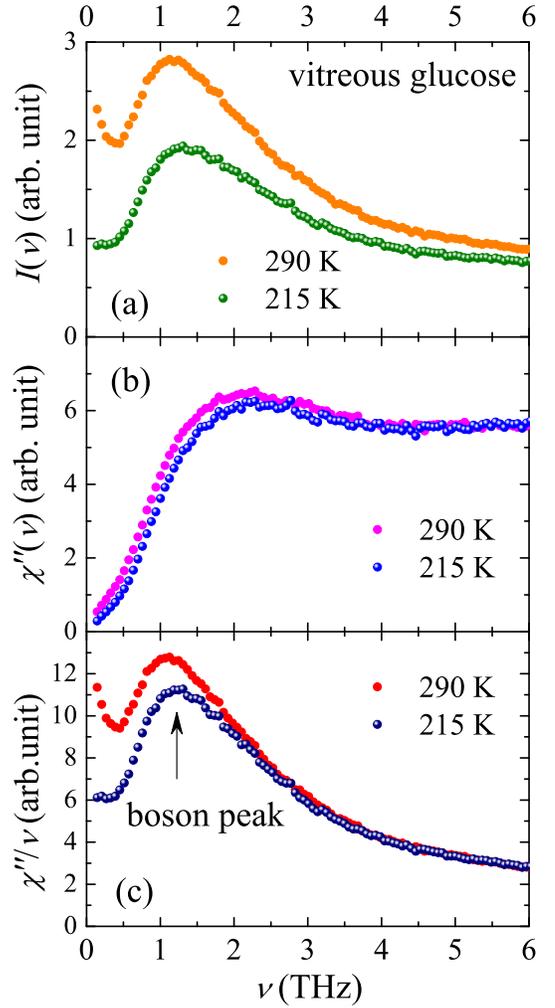}
\caption{(Color online) (a) Raman intensity spectra of the vitreous glucose on heating process. (b) The imaginary part of the Raman susceptibility $\chi''(\nu)$.  (c) The reduced Raman intensity $I_{\rm red}(\nu) \propto \chi''(\nu)/\nu$.}
\label{fig3}
\end{figure}

Low-frequency Raman spectroscopy is well known as a typical method to detect the BP \cite{Kojima1993, Surovtsev2002}, and we performed the LFRS on the vitreous glucose.
The obtained depolarized Raman spectra at 215 and 290~K are shown in Fig.~\ref{fig3}(a).
The measured Raman intensity $I(\nu)$ is related to the imaginary part of the Raman susceptibility $\chi''(\nu)$ and the reduced Raman intensity $I_{\rm red}(\nu)$ by the following equation: \cite{Hayes1978, Yannopoulos2006}
\begin{equation}
I_{\rm red}(\nu) \equiv \frac{I(\nu)}{\nu(n(\nu)+1)} \propto \frac{\chi''(\nu)}{\nu},
\label{eq-Ired}
\end{equation}
where $n(\nu) = ({\rm exp}(h\nu/k_{\rm B}T)-1)^{-1}$ which is the Bose-Einstein distribution function.
As shown in Figs.~\ref{fig3}(a) and \ref{fig3}(c), both the $I(\nu)$ and $I_{\rm red}(\nu)$ spectra at 290~K show the BP at about 1.1~THz which is slightly higher than the $\nu_{\rm BP-IR}$ at 290~K.
The results of the BP frequencies are listed in Table \ref{table1}.
These Raman spectra ($I(\nu)$, $\chi''(\nu)$, and $I_{\rm red}(\nu)$) are related to the VDOS in the disordered system by the following equation: \cite{Shuker1970, Galeener1978, Schirmacher2015, footnote1}
\begin{equation}
I_{\rm red}(\nu) \propto \frac{\chi''(\nu)}{\nu} = C_{\rm Raman}(\nu)\frac{g(\nu)}{\nu^{2}},
\label{eq-CRaman}
\end{equation}
where $C_{\rm Raman}(\nu)$ is the Raman light-vibration coupling coefficient.
The BP of the Raman spectra is often referred to as a peak of $I_{\rm red}(\nu)$, i.e., $\chi''(\nu)/\nu$, or the directly measured raw spectra $I(\nu)$.
Therefore, the $\chi''(\nu)$ spectra does not show a peak at the BP frequency $\nu_{\rm BP-Raman}$ at any temperature as shown in Fig.~\ref{fig3}(b), and this situation is almost the same as the $\varepsilon''(\nu)$ of the IR spectra.
Consequently, the BP in the Raman spectra is not a peak from the optical phonon-like mode, but is derived from the characteristic structure of $\chi''(\nu)$ of the glassy materials in the THz region.

We explain the reason for the appearance of the ``BP'' in the raw Raman intensity spectra $I(\nu)$, because it often seems confusing to new LFRS researchers.
For clarity, we re-express Eq.~(\ref{eq-Ired}) as follows:
\begin{equation}
I(\nu) \propto (n(\nu)+1)\chi''(\nu).
\label{eq-I}
\end{equation}
At room temperature, the BP frequency region (about 1~THz) is well below the frequency corresponding to room temperature (300~K $\sim$ 6~THz).
Hence, $n(\nu)+1$ at around the BP frequency is approximated by $\nu^{-1}$.
As a result, $I(\nu)$ becomes approximately $\chi''(\nu)/\nu$ and it is the reason for the similarity with $I_{\rm red}(\nu)$.
Therefore, the $boson$ $peak$ in $I(\nu)$ generally disappears at a sufficiently low temperature compared to the BP frequency energy, although the BP of $I_{\rm red}(\nu)$ shows almost no temperature dependence.
In other words, the peak of $I(\nu)$ is a result from the high temperature approximation for the quantum statistics of the Raman scattering processes regarding the non-resonant characteristic and universal structure of $\chi''(\nu)$ around the BP frequency.

\begin{table*}
\caption{\label{table1}Measured and evaluated parameters of the vitreous glucose at 290~K and 215~K. The BP frequency of VDOS is calculated as $\nu_{\rm BP-VDOS}=0.77 \times \nu_{\rm BP-Raman}$ (see text). The correlation length of the middle range order is calculated as $\xi = V_{\rm TA}/\nu_{\rm BP-VDOS}$.}
\begin{ruledtabular}
\begin{tabular}{ccccccc}
$T$ (K)&$\nu_{\rm BP-IR}$ (THz)&$\nu_{\rm BP-Raman}$ (THz)&$\nu_{\rm BP-VDOS}$ (THz)
&$V_{\rm LA}$ (m/s)&$V_{\rm TA}$ (m/s)&$\xi$ (\AA)\\ \hline
 290&1.00&1.10&0.85&$3.84 \times 10^{3}$&$9.82 \times 10^{2}$&13 \\
 215&1.08&1.24&0.95&\\
\end{tabular}
\end{ruledtabular}
\end{table*}

\subsection{Relative light-vibration coupling coefficient}

The IR and Raman spectroscopies are indirect methods for accessing the VDOS, especially the BP, although conversely, they have advantages as methods for the selective detection of those states. Therefore, in order to evaluate the interaction between light and materials in the vicinity of the BP frequency, it is crucial to understand the light-vibration coupling coefficients \cite{Shuker1970, Galeener1978, Surovtsev2002, Taraskin2006}.

\begin{figure}
\includegraphics[width=7cm]{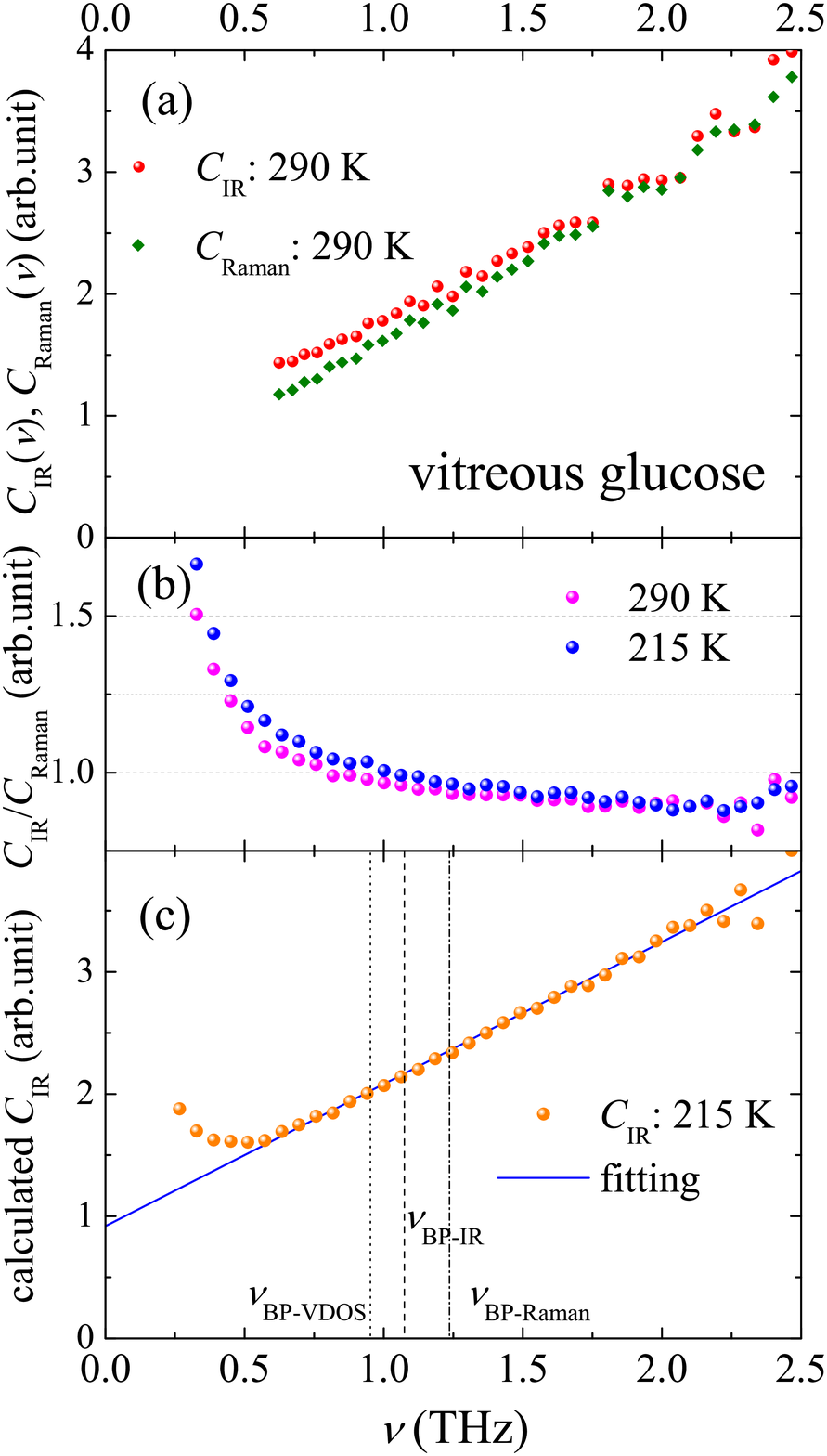}
\caption{(Color online) (a) The IR and Raman light-vibrational coupling coefficients ($C_{\rm IR}(\nu)$ and $C_{\rm Raman}(\nu)$) of the vitreous glucose at 290~K. (b) Relative light-vibrational coupling coefficient  $C_{\rm IR}(\nu)/C_{\rm Raman}(\nu) = \alpha(\nu)/(\nu \cdot \chi''(\nu))$ at 290~K and 215~K. (c) Calculated $C_{\rm IR}(\nu)$ of vitreous glucose at 215~K, assuming the $C_{\rm Raman}(\nu)$ as $A'(\nu+0.5\nu_{\rm BP-VDOS})$. A solid line is the fitting result (see text).}
\label{fig4}
\end{figure}

Figure~\ref{fig4}(a) shows the $C_{\rm IR}(\nu)$ and $C_{\rm Raman}(\nu)$ of the vitreous glucose at room temperature obtained using Eqs.~(\ref{eq-CIR}), (\ref{eq-Ired}), and the VDOS data extracted from Violini's study \cite{Violini2012}.
Note that $g(\nu)$ is determined by the hydrogenated sample in the INS study\cite{Violini2012}, and the sample is nominally identical to that of the present study. 
Both the $C_{\rm IR}(\nu)$ and $C_{\rm Raman}(\nu)$ show almost linear frequency-dependences around the BP frequency.

Taraskin $et$ $al$. proposed a model \cite{Taraskin2006} in which $C_{\rm IR}(\nu) = A + B\nu^{2}$ for below the Ioffe-Regel crossover frequency, where $A$ and $B$ are constants depending on the substances.
The frequency-independent term $A$ is due to uncorrelated static charge fluctuations caused by medium and long-range structural irregularities \cite{Taraskin2006}.
On the other hand, the quadratic frequency dependence term, $B\nu^{2}$, results from correlated charge fluctuations caused by structural disorder on the short-range (interatomic) scale, satisfying the local neutrality\cite{Taraskin2006}.
This model was evaluated for several oxide glasses by the THz-TDS \cite{Taraskin2006,Parrott2010}, however, systematic investigations for various glasses have not yet been done.
On the other hand, $C_{\rm Raman}(\nu)$ in the vicinity of the BP frequency has been well investigated  by both experimental and theoretical approaches \cite{Shuker1970, Duval1990, Duval1993, Surovtsev2002}.
Surovtsev $et$ $al$. \cite{Surovtsev2002} empirically classified the $C_{\rm Raman}(\nu)$ into two categories; type-I: $C_{\rm Raman}(\nu)=A'(\nu+0.5\nu_{\rm BP-VDOS})$, and type-II: $C_{\rm Raman}(\nu)=A''\nu$, where $A'$ and $A''$ are constants. Our obtained results of $C_{\rm Raman}(\nu)$ will be assigned as type-I and the ratio of the BP frequency of VDOS ($\nu_{\rm BP-VDOS}$) to $\nu_{\rm BP-Raman}$ becomes about 0.77 \cite{Surovtsev2002, Nakamura2015}. Therefore, the $\nu_{\rm BP-VDOS}$ is estimated to be about 0.85~THz for the vitreous glucose at room temperature, although the BP does not clearly appear in the INS spectra \cite{Violini2012} due to the contribution of the relaxation processes near the $T_{g}$.

Regarding the $C_{\rm IR}(\nu)$ at 290~K shown in Fig. \ref{fig4}(a), our result shows the linear frequency dependence even below $\nu_{\rm BP-IR}$ and it seems to deviate from Taraskin's model.
However, because room temperature is near the $T_{g}$ of the glucose, the spectrum of $\varepsilon(\nu)$ is a superposition of the VDOS peak and the excess wings \cite{Kaminski2006}.
Taraskin's model does not account for the relaxation modes.
Although the dominant contribution is the VDOS peak at 290~K as shown in Figs. \ref{fig2} and \ref{fig3}(b), the $C_{\rm IR}(\nu)$ should be evaluated at a sufficiently low temperature below the $T_{g}$.
The recent THz-TDS and LFRS can easily obtain high quality spectra and the measurement time has been decreasing with the developing detection techniques, for instance, the asynchronous optical sampling system for the fast scan THz-TDS \cite{Mori2014IOP,Kojima2014IOP}, and the CCD sensor and ultra-narrowband notch filter for LFRS \cite{Fujii2016}.
They enable the systematic study of the coupling coefficients, as shown below.

We now propose a novel analytical approach to evaluate the coupling coefficients without the VDOS spectra. We divide Eq.~(\ref{eq-af2}) by Eq.~(\ref{eq-Ired}) to eliminate the $g(\nu)$ term, and obtain the following relation:
\begin{equation}
\frac{\alpha(\nu)}{\nu\cdot\chi''(\nu)} = \frac{C_{\rm IR}(\nu)}{C_{\rm Raman}(\nu)}.
\label{eq-RCC}
\end{equation}
We call $C_{\rm IR}(\nu)/C_{\rm Raman}(\nu)$ the relative light-vibration coupling coefficient (RCC), which can be obtained from the combination of $\alpha(\nu)$ and $\chi''(\nu)$, that are experimentally determined.
Figure~\ref{fig4}(b) shows the RCC of the vitreous glucose obtained from our THz-TDS and LFRS data using Eq.~(\ref{eq-RCC}).
The RCC at 215~K shows an increase toward the low frequency below the BP frequency, and the behavior indicates that $\nu_{\rm BP-IR}$ is lower than $\nu_{\rm BP-Raman}$.

In this context, we attempted to calculate $C_{\rm IR}(\nu) = \alpha(\nu)/(\nu\cdot\chi''(\nu))\cdot C_{\rm Raman}(\nu)$ assuming the specific form of $C_{\rm Raman}(\nu)=A'(\nu+0.5\nu_{\rm BP-VDOS})$ \cite{Surovtsev2002}, and the results are shown in Fig.~\ref{fig4}(c).
The calculated $C_{\rm IR}(\nu)$ show a linearity the same as the room temperature results and the behavior deviates from the $A + B\nu^{2}$ within the experimental error.
The curling of the calculated $C_{\rm IR}(\nu)$ below 0.5 THz is due to the relaxation process as pointed out in Fig.~\ref{fig2}.
The $C_{\rm IR}(\nu)$ of 215~K is well fitted by a linear function of $\nu + 0.83\nu_{\rm BP-VDOS}$ between 0.57 and 2.1~THz.
With careful consideration about the results at both room temperature and 215~K, we concluded that the $C_{\rm IR}(\nu)$ of the vitreous glucose has a linearity even below $\nu_{\rm BP-VDOS}$, showing a deviation from Taraskin's model.

We speculate about the possible origin of the deviation.
The linearity means a discrepancy with the second term $B\nu^{2}$ of the model.
However, to derive the second term, the model assumes a simple dispersion relation of  $\omega = Q$, and the square of the relation leads to the squared frequency dependent term \cite{Taraskin2006}.
Therefore, there might be a possibility that a more realistic dispersion relation including the damping of the TA mode, especially in the vicinity of the BP,  produces the linear frequency dependence of $C_{\rm IR}(\nu)$.
As another possibility, a breakdown of the approximation formula of $\alpha(\nu) = C_{\rm IR}(\nu)g(\nu)$ would be possible; see the footnote\cite{footnote1}.

We also propose a possible category of the $C_{\rm IR}(\nu)$ having a linearity.
In the studies of the vitreous sorbitol \cite{Ruta2012, Sibik2016},  which is one of the hydrogen-bonded glass-formers, the discrepancy between $\nu_{\rm BP-VDOS}$ (1.1~THz) \cite{Ruta2012} and $\nu_{\rm BP-IR}$ (1.5~THz) \cite{Sibik2016} has been pointed out \cite{Sibik2016}.
The results suggest that the  $C_{\rm IR}(\nu)$ of the sorbitol glass might be categorized in the same group as the vitreous glucose.
Similar tendencies are also found in other polyhydric alcohols \cite{Sibik2014, Kojima1993, Yamamuro2000}.
In the vitreous glycerol\cite{Sibik2014, Kojima1993, Yamamuro2000}, $\nu_{\rm BP-IR}/\nu_{\rm BP-VDOS} \sim 1.5$, and $\nu_{\rm BP-IR}/\nu_{\rm BP-Raman} \sim 1.0$.
This indicates that these polyhydric alcohols \cite{Sibik2014} have a linear frequency dependence in the $C_{\rm IR}(\nu)$.
On the other hand, the pharmaceutical IMC glass, which is also a hydrogen-bonded organic glass, shows a significant discrepancy between the  $\nu_{\rm BP-IR}$\cite{Shibata2015, Kojima2015} and $\nu_{\rm BP-Raman}$ \cite{Shibata2015, Kojima2015, Hedoux2009}, and $\nu_{\rm BP-IR}/\nu_{\rm BP-Raman} \sim 0.6$\cite{Shibata2015, Kojima2015, Hedoux2009}.
The different tendency of the ratio from the polyhydric alcohols suggests that the $C_{\rm IR}(\nu)$ of IMC will obey Taraskin's model.
Taking into account these results, the number of intermolecular hydrogen bonds might affect the frequency dependence of the $C_{\rm IR}(\nu)$.
Although we propose possible origins and category of the linear $C_{\rm IR}(\nu)$, these issues will remain open questions.

Following the behavior of the obtained $C_{\rm IR}(\nu)$, we reconsidered the resonance-like behavior of the $\varepsilon(\nu)$ below the BP frequency shown in Fig.~\ref{fig1}.
Below the BP frequency, which corresponds to the phonon regime\cite{Nakayama2002}, the constant term of $C_{\rm IR}(\nu)$ is dominant and it produces the scaling relation between $g(\nu)$ and $\alpha(\nu)$, resulting in the linearity for $\varepsilon''(\nu)$.
On the other hand, above the BP frequency, the dominant part of the $C_{\rm IR}(\nu)$ changes to a linear dispersion resulting in the similarity between $g(\nu)$ and $\varepsilon''(\nu)$, where a bump exists in the $\varepsilon''(\nu)$.
Therefore, the $\varepsilon''(\nu)$ universally has an inflection point at around the BP frequency, showing a crossover from a negative to positive value of the second derivative of $\varepsilon''(\nu)$ toward high frequency, and the inflection point results in a peak in the  $\varepsilon''(\nu)/\nu$ or  $\alpha(\nu)/\nu^{2}$ spectra.

Consequently, the absorption band in the THz region derived from the acoustic phonon regime and the bump of $g(\nu)$ around the BP produces the resonant-like behavior in the complex response function for the glass system.
These processes occur only when the system loses its translational symmetry, therefore, such a behavior is not observed in the crystal, but appears in the glass system as the universal feature of the THz region.

\begin{figure}[t]
\includegraphics[width=8cm]{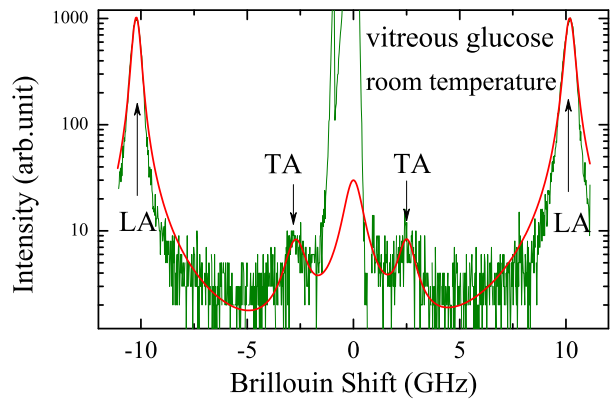}
\caption{(Color online) Brillouin scattering spectrum of the vitreous glucose with 90$^\circ$ scattering geometry at room temperature. LA denotes the longitudinal acoustic (LA) mode. The solid line shows fitting curve which consists of superposition of Voigt functions.}
\label{fig5}
\end{figure}

\begin{figure*}
\includegraphics[width=14cm]{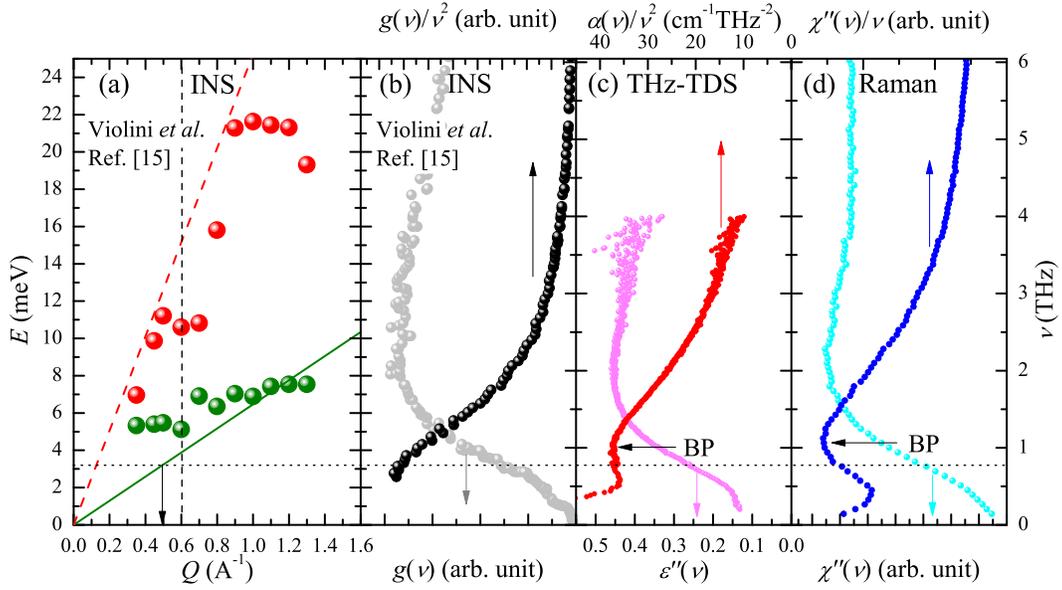}
\caption{(Color online) (a) Dispersion relations of $L$- and $H$-modes of the deutrated vitreous glucose \cite{Violini2012}. Data are depicted from Ref. [15]. The vertical dashed line indicates the position of the first peak of the static structure factor $S(Q)$ \cite{Violini2012}. (b) Vibrational density of states $g(\nu)$ and $g(\nu)/\nu^{2}$ of the hydrogenated vitreous glucose \cite{Violini2012}. Data are depicted from Ref. [15]. (c) Imaginary part of the complex dielectric constant $\varepsilon''(\nu)$ and $\alpha(\nu)/\nu^{2}$ of vitreous glucose at room temperature. (d) Imaginary part of the Raman susceptibility $\chi''(\nu)$ and $\chi''(\nu)/\nu$ of vitreous glucose at room temperature.}
\label{fig6}
\end{figure*}

\subsection{Correlation length of medium range order and pseudo Brillouin zone}

Finally, we discuss the medium range order (MRO) and pseudo Brillouin zone ($p$-BZ) of the vitreous glucose \cite{Violini2012}. We determined the sound velocities of the transverse ($V_{\rm TA}$) and  longitudinal ($V_{\rm LA}$) acoustic phonons by Brillouin light scattering (BLS) spectroscopy with a 90$^\circ$ scattering geometry configuration. Figure~\ref{fig5} shows the obtained Brillouin spectrum of the vitreous glucose at room temperature, and it results in the $V_{\rm TA}$ of $9.82 \times 10^{2}$~m/s and the $V_{\rm LA}$ of $3.84 \times 10^{3}$~m/s.
To assess the correlation length of the MRO ($\xi$), we employed the relation of $\xi = V_{\rm TA}/\nu_{\rm BP-VDOS}$, which is often used in the integrated spectroscopy of LFRS and BLS for glassy materials \cite{Duval1990, Surovtsev2002}. The calculated $\xi$ becomes 13~\AA. Assuming the shape of the MRO as cubic, the volume of the MRO has been calculated to be $\xi^{3} = 2.0 \times 10^{3}$~\AA$^3$ which is about 2.7 times greater than that of the crystalline $\alpha$- or $\beta$-D-glucose \cite{Brown1979, Chu1968}, and the MRO contains about 11 glucose molecules \cite{Parks1928}.

In this context, it is important to compare our results with the INS work done by Violini $et$ $al$. \cite{Violini2012} . Figures \ref{fig6}(a) and \ref{fig6}(b) show the dispersion relation and VDOS, respectively, and the data are extracted from Ref.~[15]. In the dispersion relation, a low lying flattened mode, called the $L$-mode around $5-7$~meV ($1.2-1.7$~THz), and a high frequency mode as the $H$-mode with a maximum of about 22~meV (5.5~THz), are observed.
The vertical dashed line indicates the position of the first peak of the static structure factor $S(Q)$ \cite{Violini2012}.
Note that Violini $et$ $al$. did not assign it as the first sharp diffraction peak in their paper \cite{Violini2012}.
As shown in Fig.~\ref{fig6}(b), $g(\nu)$ shows a broad and shoulder-like peak at about 7~meV, which might mainly reflect the structure of the flattened $L$-mode \cite{Violini2012}.
On the other hand, $g(\nu)/\nu^{2}$ shows only a slope at 7~meV, and a blurred peak-like structure at around 3~meV (0.7~THz) corresponding to $\nu_{\rm BP-VDOS}$ was evaluated from our results (horizontal dotted line).
Regarding the lowering of the $\nu_{\rm BP-VDOS}$ than the $L$-mode frequency, it is naturally understood considering the following.
Generally, the longitudinal current spectra $C^{\rm L}(Q, \omega)$ is proportional to the dynamic structure factor multiplied by the squared angular frequency $\omega^{2}S(Q, \omega)$ \cite{Pilla2004, Schirmacher2015}, and the resonant frequency of the DHO model corresponds to the peak frequency of $C^{\rm L}(Q, \omega)$.
When the damping of the DHO mode is small, the peak frequencies of $C^{\rm L}(Q, \omega)$ and $S(Q, \omega)$ match.
However, large damping results in the lowering the peak frequency of the $S(Q, \omega)$ more than that of the $C^{\rm L}(Q, \omega)$.
The $g(\nu)$ and $g(\nu)/\nu^{2}$ correspond to the $Q$-integral of the $C^{\rm L}(Q, \omega)$ and $S(Q, \omega)$, respectively.
Therefore, the $\nu_{\rm BP-VDOS}$ becomes lower than the $L$-mode frequency due to the large damping.
Similar results were observed in the recent INS study about glassy SiSe$_{2}$ \cite{Zanatta2013}.

We present our IR and Raman results in Figs.~\ref{fig6}(c) and (d), and the $V_{\rm TA}$ and $V_{\rm LA}$ in Fig.~\ref{fig6}(a) as dashed and solid linear lines, respectively. The similarity with $g(\nu)$ and $\varepsilon''(\nu)$ or $\chi''(\nu)$ is clear,  and the shoulder-like peak around $7-8$~meV for both $\varepsilon''(\nu)$ or $\chi''(\nu)$ might have originated from the entire structure of the VDOS, especially from the flat $L$-mode in the vitreous glucose.

Regarding the size of $p$-BZ, we have estimated $Q_{\rm BP-VDOS} = 2\pi/\xi_{\rm BP-VDOS}$, as 0.50~\AA$^{-1}$, as indicated by the arrow in Fig.~\ref{fig6}(a).
However, the extrapolated dispersion curve of the TA mode from our results of the BLS (a solid line in Fig.~\ref{fig6}(a)) shows an apparent discontinuity with the flattened $L$-mode, although the LA mode from the results of the BLS and the $H$-mode are in good agreement.
Moreover, the extrapolated TA mode from our BLS result reaches the $L$-mode at about 1.0~\AA$^{-1}$.
One possibility to explain the observed $V_{\rm TA}$ and $L$-mode behavior might be a hybridization of the TA mode and the flat $L$-mode.
It will result in a large broadening of the TA mode, and therefore, the TA mode might possibly not have appeared as a peak in the INS study \cite{Violini2012}.
It is worth noting that the momentum transfer of the Ioffe-Regel crossover of the $H$-mode is about 1.0~\AA$^{-1}$, where the $H$-mode shows a maximum \cite{Violini2012}.
Consequently, if above speculation is correct, our observation of the $V_{\rm TA}$ by BLS seems to be totally consistent with both the $L$-and $H$-modes.
In addition, the $H$-mode shows an anomalous kink at around 0.6~\AA$^{-1}$ with a frequency of about 2.6~THz.
The crystalline $\alpha$-D-glucose shows strong optic modes in the $\Gamma$ point at 1.4 and 2.6~THz \cite{Walther2003}, and these frequencies correspond to the flattened $L$-mode and the anomaly point of the $H$-mode, respectively.
A more detailed comparison with the crystalline system and further investigation of the dynamic structure of the lower $Q$ region will be required.

\section{\label{sec:level1}Conclusions}

We performed the THz-TDS, LFRS, and BLS on vitreous glucose to investigate the universal BP dynamics.
In the spectra of $\alpha(\nu)/\nu^{2}$, the BP is clearly observed at around 1.1~THz, and it is shown that the THz-TDS is suitable to detect the BP.
The complex dielectric constants show a universal resonant-like behavior below $\nu_{\rm BP-IR}$, resulting from the combination of the interaction between the light and acoustic phonon in the disordered material and the bump of the VDOS around the BP frequency.
It is re-noted that the BP in the Raman spectra is not a peak in $\chi''(\nu)$, but in $\chi''(\nu)/\nu$.
The relative light-vibration coupling coefficient, $C_{\rm IR}(\nu)/C_{\rm Raman}(\nu)$, is proposed, which is obtainable from the experimental spectra of $\alpha(\nu)$ and $\chi''(\nu)$ without $g(\nu)$.
The $C_{\rm IR}(\nu)$ of the vitreous glucose was calculated assuming the specific form of $C_{\rm Raman}(\nu)$ being proportional to $\nu+0.5\nu_{\rm BP-VDOS}$ \cite{Surovtsev2002}, and it shows a linearity and deviates from Taraskin's model \cite{Taraskin2006}.
Although we pointed out the several possibilities of the origin and categorization of the linear $C_{\rm IR}(\nu)$, these problems will be open questions and further investigations are required for not only the vitreous glucose, but also the whole glass systems.
From the BLS results, we found a discontinuity between the measured TA mode and the flattened $L$-mode \cite{Violini2012}, which suggests the coupling of these modes.
Based on a comparison between our results and the INS study \cite{Violini2012}, the behaviors of the $\varepsilon''(\nu)$, $\chi''(\nu)$, and $g(\nu)$ are totally consistent.
The integrated THz-band spectroscopy, i.e., the combination of THz-TDS, LFRS, and BLS is effective for the systematic investigation of the universal THz dynamics of a glass system as a complementary method of the INS and IXS.
The re-recognition of the detection of the BP by far-infrared spectroscopy will become a significant key to elucidate the nature of the boson peak dynamics.

\begin{acknowledgments}
The authors are thankful to Yu Matsuda for the crucial idea and advice about the BP detection by THz-TDS. T. M. is thankful to Masato Matsuura for the kind advice about interpretation of the INS experiment. T. M. is thankful to Shinji Kohara for the fruitful discussion about interpretation of the $S(Q)$. T. M. is thankful to Takanari Kashiwagi for the heartful discussion about the making manuscript. The authors are thankful to Tomohiko Shibata, Shota Koda, and Yusuke Hashimoto for the technical advice about the sample preparation and fruitful discussions. The authors are thankful to the technical support for the broadband THz-TDS measurements by the Advantest Corporation. This work was partially supported by JSPS KAKENHI Grant Numbers 24740194 and 26287067, Nippon Sheet Glass Foundation for Materials Science and Engineering, the Murata Science Foundation, and the National Research Foundation of Korea (NRF) grant funded by the Korea government (MSIP) (NRF-2016R1A2B4012646).
\end{acknowledgments}

\bibliography{References}

\end{document}